\documentclass[11pt,preprintnumbers,titlepage,nofootinbib]{revtex4-2}
\usepackage[latin9]{inputenc}
\usepackage{color}
\usepackage{array}
\usepackage{multirow}
\usepackage{amsmath}
\usepackage{amssymb}
\usepackage{graphicx}
\usepackage[unicode=true,
 bookmarks=false,
 breaklinks=false,pdfborder={0 0 1},colorlinks=false]
 {hyperref}
\hypersetup{
 colorlinks,linkcolor=blue,citecolor=blue,urlcolor=blue}

\makeatletter

\providecommand{\tabularnewline}{\\}

\usepackage{wasysym}
\usepackage{array}
\usepackage{multirow}
\usepackage{wasysym}
\usepackage{wasysym}
\usepackage{amsfonts}
\usepackage{subfigure}
\usepackage{cleveref}
\usepackage{listings}
\usepackage{xcolor}
\usepackage{array}
\usepackage{url}
\usepackage{color}
\usepackage{subfigure}
\providecommand{\tabularnewline}{\\}
\@ifundefined{textcolor}{}{
\definecolor{BLACK}{gray}{0}
\definecolor{WHITE}{gray}{1}
\definecolor{RED}{rgb}{1,0,0}
\definecolor{GREEN}{rgb}{0,1,0}
\definecolor{BLUE}{rgb}{0,0,1}
\definecolor{CYAN}{cmyk}{1,0,0,0}
\definecolor{MAGENTA}{cmyk}{0,1,0,0}
\definecolor{YELLOW}{cmyk}{0,0,1,0}
\definecolor{ballblue}{rgb}{0.13, 0.67, 0.8}
\definecolor{bleudefrance}{rgb}{0.19, 0.55, 0.91}
\definecolor{blue(ncs)}{rgb}{0.0, 0.53, 0.74}
\definecolor{darkpastelgreen}{rgb}{0.01, 0.75, 0.24}
\definecolor{darkspringgreen}{rgb}{0.09, 0.45, 0.27}
\definecolor{denim}{rgb}{0.08, 0.38, 0.74}
\definecolor{electricviolet}{rgb}{0.56, 0.0, 1.0}
}

\makeatother

\begin{document}
\title{Optical Appearance of Scalarized Kerr-Newman Black Holes
with Multiple Light Rings}
\author{Yiqian Chen$^{a}$}
\email{chenyiqian@ucas.ac.cn}

\author{Li Li$^{a,b,c}$}
\email{liliphy@itp.ac.cn}

\author{Peng Wang$^{d}$}
\email{pengw@scu.edu.cn}

\affiliation{$^{a}$School of Fundamental Physics and Mathematical Sciences, Hangzhou
Institute for Advanced Study, University of Chinese Academy of Sciences,
Hangzhou, 310024, China}
\affiliation{$^{b}$Institute of Theoretical Physics, Chinese Academy of Sciences,
Beijing, 100190, China}
\affiliation{$^{c}$School of Physical Sciences, University of Chinese Academy
of Sciences, Beijing, 100049, China}
\affiliation{$^{d}$College of Physics, Sichuan
University, Chengdu, 610064, China}
\begin{abstract}
This study investigates the optical appearance of rotating scalarized
Kerr-Newman black holes in the Einstein-Maxwell-scalar theory with
exponential coupling. By analyzing equatorial null geodesics, these
black holes are classified into six types according to the number and properties of their light rings. Combining slow-rotation analysis with full numerical ray tracing, we investigate images of black holes illuminated by a geometrically and optically thin accretion disk. Unlike the Kerr case, where the image is typically governed by a single outer photon shell and a single critical curve, scalarized Kerr-Newman black holes can develop an additional inner photon shell outside the event horizon. This extra shell gives rise to an inner critical curve inside the usual outer one, which may be absent, partial, or closed depending on the black hole parameters and the observer's inclination. Moreover, it generates new higher-order images in the region between the two critical curves, some of which exhibit crescent-like morphologies distinct from the nearly circular higher-order images familiar from Kerr black holes. These features enrich the optical appearance of scalarized black holes and could serve as distinctive observational signatures in future high-resolution observations.
\end{abstract}
\maketitle
\tableofcontents{}

{}

\section{Introduction}

\label{sec:Introduction}

The Event Horizon Telescope (EHT) collaboration has substantially
enhanced our understanding of black holes through horizon-scale images
of M87{*} and Sagittarius A{*}, revealing a bright ring surrounding
a central intensity depression~\cite{Akiyama:2019cqa,Akiyama:2019bqs,Akiyama:2019fyp,EventHorizonTelescope:2022xnr,EventHorizonTelescope:2022wok}.
This characteristic feature arises from synchrotron emission produced
by relativistic plasma near the event horizon, thereby opening a new
era for testing general relativity in the strong-field regime. It
has been shown that the bright ring is composed of a sequence of nested
images of the accreting plasma, commonly referred to as the photon
ring, which asymptotically approaches the boundary of the central
shadow, known as the critical curve~\cite{Gralla:2019xty}. The prominent photon ring structure is closely related to strong light deflection near unstable bound photon orbits, which form a photon sphere in Schwarzschild spacetime or a photon shell in Kerr spacetime. Therefore, the photon ring and the critical curve serve as direct probes of the background geometry, largely independent of the details of the astrophysical
emission model.

Comparing the EHT observations with predictions from general relativistic
magnetohydrodynamic (GRMHD) simulations, the Kerr hypothesis has been
shown to be broadly consistent with the measured brightness distribution
and ring diameter, while still allowing room for potential deviations
within current observational uncertainties. Consequently, EHT measurements
provide powerful tools for constraining alternative models that depart
from the Kerr geometry, offering new insights into the near-horizon
environment and possible extensions of general relativity~\cite{Li:2013jra,Tsukamoto:2014tja,Kumar:2018ple,Bambi:2019tjh,Afrin:2022ztr,Wang:2025fmz}.
Related studies have also been extended to additional observables,
such as polarization signatures~\cite{EventHorizonTelescope:2021btj,Gelles:2021kti,Qin:2021xvx,Delijski:2022jjj,Qin:2022kaf,Lee:2022rtg,Zhu:2022amy,Liu:2022ruc,Hu:2022sej,Qin:2023nog,Deliyski:2023gik},
relativistic jets~\cite{Lu:2023bbn,Papoutsis:2022kzp,Zhang:2024lsf}
and flares~\cite{abuter2018detection,GRAVITY:2020lpa,Wielgus:2022heh,Huang:2024wpj},
further enriching our understanding of the astrophysical and gravitational
processes occurring near compact objects. 

Due to the high computational cost of GRMHD simulations, simplified
accretion disk models are often employed to capture the essential
features of black hole images. Using such models, black hole images
and associated optical phenomena have been extensively investigated
across a wide range of theoretical frameworks, including string inspired
black holes~\cite{Amarilla:2011fx,Guo:2019lur,Zhu:2019ura,Kumar:2020hgm},
fuzzball~\cite{Bacchini:2021fig}, exotic ultra-compact objects~\cite{Cunha:2017wao,Cunha:2018acu,Shaikh:2018oul,Shaikh:2019itn,Shaikh:2019jfr,Wielgus:2020uqz,Peng:2021osd,Huang:2024bbs},
naked singularities~\cite{Joshi:2020tlq,Dey:2020bgo,Chen:2023trn,Chen:2023knf,Deliyski:2024wmt},
and modified gravity theories~\cite{Amarilla:2010zq,Ayzenberg:2018jip,Wang:2018prk,Ma:2019ybz,Guo:2020zmf,Wei:2020ght,Zeng:2020dco,Addazi:2021pty,Wang:2022yvi,Liu:2024lve}. A promising theoretical framework is provided by Einstein-Maxwell-Scalar
(EMS) theories~\cite{Herdeiro:2018wub}. In these theories, a scalar
field $\phi$ is non-minimally coupled to the electromagnetic fields.
Such coupling can trigger tachyonic instabilities in charged black
hole backgrounds, resulting in a new class of solutions known as scalarized
Reissner-Nordstr\"om (RN) black holes. This discovery has spurred extensive
research within the EMS framework in asymptotically flat spacetime, including various extensions of interaction terms~\cite{Fernandes:2019rez,Fernandes:2019kmh,Blazquez-Salcedo:2020nhs}, massive and self-interacting scalar fields~\cite{Zou:2019bpt,Fernandes:2020gay}, horizonless reflecting stars~\cite{Peng:2019cmm}, quasinormal modes and stability analysis~\cite{Myung:2018vug,Myung:2018jvi,Myung:2019oua,Zou:2020zxq,LuisBlazquez-Salcedo:2020rqp,Myung:2020etf,Mai:2020sac}, higher dimensional scalar-tensor models~\cite{Astefanesei:2020qxk}, dynamical scalarization and descalarization~\cite{Zhang:2021nnn,Zhang:2022cmu,Jiang:2023yyn}, critical phenomena inside scalarized black holes~\cite{LiLi:2025qgo}, and rotating scalarized black hole solutions~\cite{Guo:2023mda}.

Intriguingly, scalarized RN black holes can possess multiple photon
spheres outside the event horizon within certain parameter ranges
\cite{Gan:2021pwu}. This distinctive property has motivated intensive
investigations into their optical appearance, including accretion
disks~\cite{Gan:2021pwu,Gan:2021xdl,Chen:2023qic}, luminous celestial
spheres~\cite{Guo:2022muy}, hot spots~\cite{Chen:2024ilc} and infalling
emitters~\cite{Chen:2022qrw}. These studies revealed that the presence
of an additional photon sphere can create extra higher-order images,
significantly enhance the observed accretion-disk flux, generate beat
patterns in the visibility amplitude, induce secondary peaks in the
light curves of orbiting hot spots, and give rise to additional flashes
from infalling emitters. 

Most existing studies have focused on spherically symmetric scalarized black holes, while the optical manifestations of their rotating counterparts, known as scalarized Kerr-Newman (KN) black holes, remain largely unexplored. A first step toward understanding their null-geodesic structure was taken in~\cite{Guo:2023mda}, which analyzed light rings on the equatorial plane and revealed a rich dependence on spin and charge. However, the optical appearance of this class of black holes has not yet been studied. In this work, we fill this gap by investigating the observational signatures produced by thin accretion disks around these rotating scalarized black holes, focusing on how the presence of multiple light rings affects the disk images. Our analysis uncovers nontrivial imprints on black hole images, which may serve as a distinctive signature of scalarized KN black holes.

The paper is organized as follows. In Section~\ref{sec:Setup}, we review the scalarized KN black hole solutions in the EMS model, analyze their geodesic structure, and outline the ray-tracing setup. Section~\ref{sec:Lensing-Images} presents a qualitative analysis of critical curves under the slow-rotation approximation, which we later validate through simulated images of thin accretion disks. We summarize and conclude in Section~\ref{sec:CONCLUSIONS}. Throughout the paper, we adopt the convention $G=c=1$.

\section{Setup}

\label{sec:Setup}

This section first presents a concise review of the scalarized KN
black hole solution within the 4-dimensional EMS model. We then review
circular null and timelike orbits in this spacetime, which give rise
to two key structures: light rings and marginally stable circular
orbits, respectively.

\subsection{Scalarized KN Black Hole}

In the EMS model, the scalar field $\phi$ is non-minimally coupled
to the electromagnetic field $A_{\mu}$ through some coupling function
$f\left(\phi\right)$, as described by the action
\begin{equation}
S=\frac{1}{16\pi}\int d^{4}x\sqrt{-g}\left[\mathcal{R}-2\partial_{\mu}\phi\partial^{\mu}\phi-f\left(\phi\right)F_{\mu\nu}F^{\mu\nu}\right],\label{eq:Action}
\end{equation}
where $\mathcal{R}$ is the Ricci scalar, and $F_{\mu\nu}=\partial_{\mu}A_{\nu}-\partial_{\nu}A_{\mu}$
is the electromagnetic field strength tensor. This study focuses on
the exponential coupling function $f\left(\phi\right)=e^{\alpha\phi^{2}}$
with $\alpha>0$, which satisfies $f\left(0\right)=1$ and $f^{\prime}\left(0\right)=0$
and allows the existence of scalar-free black holes like KN black
holes~\cite{Herdeiro:2018wub,Fernandes:2019rez}.

In the scalar-free background, a small scalar perturbation $\delta\phi$
obeys the equation
\begin{equation}
\left(\nabla_{\mu}\nabla^{\mu}-\mu_{\text{eff}}^{2}\right)\delta\phi=0,\label{eq:lineareq}
\end{equation}
where the effective mass is given by $\mu_{\text{eff}}^{2}=\alpha F_{\mu\nu}F^{\mu\nu}/2$.
For the KN background, 
\begin{align}\label{eq:Kerr}
ds^2
=&
-\left(1-\frac{2Mr-Q^2}{\Sigma}\right)dt^2
-\frac{2(2Mr-Q^2)a\sin^2\theta}{\Sigma}\,dt\,d\phi
+\frac{\Sigma}{\Delta}\,dr^2 \\ \nonumber
&+\Sigma\,d\theta^2
+\left(r^2+a^2+\frac{(2Mr-Q^2)a^2\sin^2\theta}{\Sigma}\right)\sin^2\theta\,d\phi^2,
\end{align}
with $\Sigma = r^2+a^2\cos^2\theta$ and $\Delta = r^2-2Mr+a^2+Q^2$,
the effective mass square reads
\begin{equation}
\mu_{\text{eff}}^{2}=-\frac{\alpha Q^{2}\left(r^{4}-6a^{2}r^{2}\cos^{2}\theta+a^{4}\cos^{4}\theta\right)}{\left(r^{2}+a^{2}\cos^{2}\theta\right)^{4}},
\end{equation}
where $Q$ is the black hole charge, and $a=J/M$ is the spin parameter. When $\mu_{\text{eff}}^{2}<0$, the spacetime~\eqref{eq:Kerr} exhibits tachyonic instabilities,
which can spontaneously induce a non-trivial scalar field, resulting
in the scalarized black hole solution.

The scalarized KN solutions are then obtained by solving the full equations
of motion derived from the action $\left(\ref{eq:Action}\right)$,
which comprise the Einstein, scalar, and Maxwell equations,
\begin{align}
\mathcal{R}_{\mu\nu}-\frac{1}{2}\mathcal{R} g_{\mu\nu} & =2T_{\mu\nu},\nonumber \\
\nabla_{\mu}\nabla^{\mu}\phi-\frac{\alpha}{2}\phi e^{\alpha\phi^{2}}F_{\mu\nu}F^{\mu\nu} & =0,\label{eq:EOM}\\
\nabla_{\mu}\left(e^{\alpha\phi^{2}}F^{\mu\nu}\right) & =0,\nonumber 
\end{align}
where $\mathcal{R}_{\mu\nu}$ is the Ricci tensor, and $T_{\mu\nu}$ is the
energy-momentum tensor expressed as 
\begin{equation}
T_{\mu\nu}=\partial_{\mu}\phi\partial_{\nu}\phi-\frac{1}{2}g_{\mu\nu}\partial_{\rho}\phi\partial^{\rho}\phi+e^{\alpha\phi^{2}}\left(F_{\mu\rho}{F_{\nu}}^{\rho}-\frac{1}{4}g_{\mu\nu}F_{\rho\sigma}F^{\rho\sigma}\right).
\end{equation}
To obtain rotating solutions, we assume a stationary, axisymmetric
and asymptotically-flat black hole ansatz,
\begin{align}
ds^{2} & =-e^{2F_{0}}Ndt^{2}+e^{2F_{1}}\left(\frac{dr^{2}}{N}+r^{2}d\theta^{2}\right)+e^{2F_{2}}r^{2}\sin^{2}\theta\left(d\varphi-\frac{W}{r^{2}}dt\right)^{2}\,,\nonumber \\
A_{\mu}dx^{\mu} & =\left(A_{t}-A_{\varphi}\frac{W}{r^{2}}\sin\theta\right)dt+A_{\varphi}\sin\theta d\varphi,\label{eq:ansatz}\\
\phi & =\phi\left(r,\theta\right),\nonumber 
\end{align}
where $N\equiv1-r_{H}/r$ with $r_{H}$ the radius of the event
horizon. The metric functions $F_{0}$, $F_{1}$, $F_{2}$, $W$,
$A_{t}$, $A_{\varphi}$ depend solely on the coordinates $r$ and
$\theta$. Substituting the ansatz $\left(\ref{eq:ansatz}\right)$
into the equations of motion $\left(\ref{eq:EOM}\right)$ yields a
coupled system of nonlinear partial differential equations (PDEs).
Both scalar-free and scalarized KN black holes emerge as solutions
of this system. We numerically solve these PDEs using pseudospectral
methods, applying appropriate boundary conditions for self-consistency:
regularity at the symmetric axis, flatness at spatial infinity, and
well-behaved behavior at the event horizon~\cite{Guo:2023mda}. The black hole mass $M$, charge $Q$, angular momentum $J$, electrostatic potential $\Phi$, and horizon angular velocity $\Omega_H$ are extracted from the asymptotic behavior of the fields at the event horizon and at spatial infinity,
\begin{align}
\left.A_{t}\right|_{r=r_{H}} & \sim0, & \left.W\right|_{r=r_{H}} & \sim r_{H}^{2}\Omega_{H},\\
\left.A_{t}\right|_{r=r_{\infty}} & \sim\Phi-\frac{Q}{r}, & \left.W\right|_{r=r_{\infty}} & \sim\frac{2J}{r}, & \left.e^{2F_{0}}N\right|_{r=r_{\infty}} & \sim1-\frac{2M}{r}.\nonumber 
\end{align}

Before proceeding, we briefly discuss the issue regarding the astrophysical relevance of our EMS theory. If $A_\mu$ is identified with ordinary electromagnetism, the scalarized black hole carries electric charge. Astrophysical plasmas, however, efficiently neutralize net charge, making such configurations negligible in practice. The theory then serves primarily as a toy model for studying spontaneous scalarization, a phenomenon of considerable theoretical interest. Nevertheless, the scalarized KN geometry~\eqref{eq:ansatz} admits a dual interpretation that may be more astrophysically natural. Introducing the dual field strength \(\tilde{F} = f(\phi) * F\) and rewriting the electromagnetic sector as \(f(\phi)^{-1} \tilde{F}^2\) leaves the equations of motion unchanged. This dual description favors magnetically dominated environments, which are observationally far more plausible than net electric charge. Magnetic black holes have garnered growing interest following the work of Maldacena~\cite{Maldacena:2020skw}.

Alternatively, one may interpret $A_\mu$ as a dark photon from a hidden sector. The associated dark charge does not couple to ordinary matter and thus evades plasma neutralization. In this interpretation, spontaneous scalarization provides a novel mechanism for black holes to interact with dark matter, potentially leading to observable imprints in strong gravity regions. This lies at the intersection of modified gravity, dark matter, and multimessenger astronomy. Dark photon scenarios have received much recent attention. However, any viable model must satisfy stringent constraints from gravitational waves, black hole and cosmological surveys, and particle physics experiments.

\subsection{Geodesic Structure}

The geodesics of particles in the spacetime are governed by the Hamiltonian
\begin{equation}
\mathcal{H}\equiv\frac{1}{2}g^{\mu\nu}p_{\mu}p_{\nu}, \quad \frac{dx^{\mu}}{d\lambda}=p^{\mu}\,,
\end{equation}
with $\lambda$ the affine parameter. Due to stationarity and axisymmetry, $\mathcal{H}$ is independent
of the coordinates $t$ and $\varphi$. Consequently, two conserved
quantities exist along the geodesic:
\begin{align}
E & \equiv-p_{t}=-g_{tt}\dot{t}-g_{t\varphi}\dot{\varphi},\nonumber \\
L_{z} & \equiv p_{\varphi}=g_{t\varphi}\dot{t}+g_{\varphi\varphi}\dot{\varphi},\label{eq:EandL}
\end{align}
where dots denote differentiation with respect to $\lambda$. 

For a massless particle, $E$ and $L_{z}$ represent its energy and
angular momentum, respectively. By isolating the positive kinetic
energy term from the Hamiltonian $\mathcal{H}=0$, one can introduce
a rescaled effective potential~\cite{Cunha:2016bjh}
\begin{align}
V & \equiv g_{tt}\left(b-h_{+}\right)\left(b-h_{-}\right)\geq0,\label{eq:=000020rescaled=000020V}
\end{align}
where $b\equiv L_{z}/E$ is the impact parameter, and
\begin{equation}
h_{\pm}=\frac{-g_{t\varphi}\mp\sqrt{D}}{g_{tt}},\quad D\equiv g_{t\varphi}^{2}-g_{tt}g_{\varphi\varphi}.
\end{equation}
Outside the ergoregion $\left(g_{tt}<0\right)$, the two functions
satisfy $h_{-}<0<h_{+}$. Photon motion with a given impact parameter
$b$ is therefore allowed only in the region $h_{-}<b<h_{+}$. The
curves $h_{+}=b$ and $h_{-}=b$ therefore define the boundaries of
allowed regions for prograde and retrograde photon trajectories, respectively.
Accordingly, $h_{+}$ and $h_{-}$ are also commonly regarded as effective
potentials. Inside the ergoregion $\left(g_{tt}>0\right)$, the inequality
reverses, leading to $h_{+}<h_{-}$. Photon motion is then permitted
in regions $b>h_{-}$ or $b<h_{+}$. In this paper, we mainly consider black holes with relatively small spin. In this regime, the ergoregion is confined to a narrow region near the event horizon. We therefore restrict the following effective potential analysis to null geodesics outside the ergoregion. This is sufficient for the main optical features studied in this work. A full analysis of geodesics inside the ergoregion is left for future work.

A light ring refers to a closed circular null orbit on the equatorial
plane. It satisfies $p_{r}=p_{\theta}=\dot{p_{r}}=0$, which is equivalent
to the conditions $V=\partial_{r}V=0$. The condition $V=0$ yields
$b=h_{+}$ or $b=h_{-}$. Differentiating Eq. $\left(\ref{eq:=000020rescaled=000020V}\right)$,
one obtains
\begin{equation}
\partial_{r}V=\pm g_{tt}\left(h_{+}-h_{-}\right)\partial_{r}h_{\pm}.
\end{equation}
Since $h_{-}<0<h_{+}$, the condition $\partial_{r}V=0$ reduces to
$\partial_{r}h_{\pm}=0$. Moreover, an unstable (stable) light ring
satisfies $\partial_{r}^{2}V>0$ ($\partial_{r}^{2}V<0$), corresponding
to $\pm\partial_{r}^{2}h_{\pm}>0$ ($\pm\partial_{r}^{2}h_{\pm}<0$).
For scalarized KN black holes, the number and stability of light rings
have been systematically analyzed in prior work. Following the classification
established in~\cite{Guo:2023mda}, scalarized KN black hole solutions
can be categorized into six types as summarized
in Table~\ref{Table:=000020classification}. In what follows,
we mainly focus on the parameter region with relatively large charge
and relatively small spin, where the black holes exhibit multiple
light rings. By contrast, in other regions of the parameter space,
the number of light rings is the same as in Kerr black holes and therefore
is not expected to lead to significant observational differences.

\begin{table}
\begin{tabular}{ccccccc}
\hline 
\multirow{3}{*}{Type} & \multicolumn{3}{c}{Prograde light ring} & \multicolumn{3}{c}{Retrograde light ring}\tabularnewline
\cline{2-7}
 & \multirow{2}{*}{Unstable} & \multirow{2}{*}{Stable} & \multirow{2}{*}{Visible inner} & \multirow{2}{*}{Unstable} & \multirow{2}{*}{Stable} & \multirow{2}{*}{Visible inner}\tabularnewline
 &  &  &  &  &  & \tabularnewline
\hline 
I & 1 & 0 & / & 1 & 0 & /\tabularnewline
\hline 
II1 & 2 & 1 & NO & 1 & 0 & /\tabularnewline
II2 & 2 & 1 & YES & 1 & 0 & /\tabularnewline
\hline 
III1 & 2 & 1 & NO & 2 & 1 & NO\tabularnewline
III2 & 2 & 1 & YES & 2 & 1 & NO\tabularnewline
III3 & 2 & 1 & YES & 2 & 1 & YES\tabularnewline
\hline 
\end{tabular}\caption{Classification of scalarized KN black holes by the number and visibility
of light rings, following the scheme established in~\cite{Guo:2023mda}.
The header \textquotedblleft Visible inner\textquotedblright{} denotes
whether near critical photons emitted from the inner light rings are
able to escape past the outer light rings on the equatorial plane.}

\label{Table:=000020classification}
\end{table}

To construct an accretion disk model, we study the marginally stable
circular orbit (MSCO) of massive particles in scalarized KN black
holes. For a massive particle, $E$ and $L_{z}$ represent its energy
and angular momentum per unit mass. On the equatorial plane, the timelike
geodesic equation is simplified as
\begin{align}
g_{rr}\dot{r}^{2}+V_{m}\left(r,E,L_{z}\right) & =0,
\end{align}
where
\begin{equation}
V_{m}\left(r,E,L_{z}\right)=1+g^{tt}E^{2}+g^{\varphi\varphi}L_{z}^{2}-2g^{t\varphi}EL_{z},
\end{equation}
is the effective potential for equatorial timelike particles. Solving
the conditions for a circular orbit $V_{m}=\partial_{r}V_{m}=0$ alongside
the marginal stability condition $\partial_{r}^{2}V_{m}=0$ yields
the location, energy and angular momentum per unit mass of the MSCO. 

In the Kerr spacetime, two MSCOs exist, corresponding to the prograde
and retrograde innermost stable circular orbits (ISCOs), and stable
circular orbits extend outward from the ISCO to infinity.\textcolor{black}{{}
For convenience, we will continue to refer to the prograde MSCO that
extends out to infinity as the ISCO. }However, in certain non-Kerr
spacetimes, additional MSCOs may appear. For example, in the Manko-Novikov
metric an island of stable equatorial circular orbits can arise in
the region $r<r_{\text{ISCO}}$~\cite{Bambi:2013eb}. A similar stable-orbit
island also emerges in  some of the scalarized KN black holes discussed here. In our thin disk model, the main disk is defined by the outer branch of stable circular orbits. Matter inside the ISCO then plunges inward with the specific energy and angular momentum fixed at the ISCO. For solutions with a stable-orbit island, such plunging trajectories do not populate the inner island. A proper treatment of emission from that region would therefore require additional assumptions. For the slowly rotating solutions considered here, however, the island is confined to a very narrow radial interval, so its effect on the optical appearance is expected to be small. We therefore neglect this stable-orbit island in the present work.

\subsection{Ray-tracing Implementation}

To obtain images of scalarized KN black holes, we trace light rays
from the observer to the emitting source by numerically integrating
the geodesic equations
\begin{equation}
\frac{dx^{\mu}}{d\lambda}=p^{\mu},\quad\frac{dp^{\mu}}{d\lambda}=-\Gamma_{\rho\sigma}^{\mu}p^{\rho}p^{\sigma},
\end{equation}
where $\Gamma_{\rho\sigma}^{\mu}$ is the Christoffel symbol. In rotating
spacetimes, we utilize a zero-angular-momentum observer (ZAMO) located
at $x_{o}^{\mu}=\left(t_{o},r_{o},\theta_{o},\varphi_{o}\right)$. The zero-angular-momentum condition $L_{z}=0$ determines the angular velocity $\omega=-g_{t\varphi}\left(r_{o},\theta_{o}\right)/g_{\varphi\varphi}\left(r_{o},\theta_{o}\right)$
and the ZAMO four-velocity
\begin{equation}
u=\frac{1}{k}\left(\partial_{t}+\omega\partial_{\varphi}\right),\quad k=\left.\sqrt{g_{t\varphi}^{2}/g_{\varphi\varphi}-g_{tt}}\right|_{x^{\mu}=x_{o}^{\mu}},
\end{equation}
where $k$ is the lapse. The associated orthonormal tetrad is
\begin{equation}
e_{\left(t\right)}=u,\quad e_{\left(r\right)}=\frac{1}{\sqrt{g_{rr}\left(r_{o},\theta_{o}\right)}}\partial_{r},\quad e_{\left(\theta\right)}=\frac{1}{\sqrt{g_{\theta\theta}\left(r_{o},\theta_{o}\right)}}\partial_{\theta},\quad e_{\left(\varphi\right)}=\frac{1}{\sqrt{g_{\varphi\varphi}\left(r_{o},\theta_{o}\right)}}\partial_{\varphi}.
\end{equation}
A photon's momentum in the local frame is expressed as $p^{\mu}=e_{\;\left(a\right)}^{\mu}p^{\left(a\right)}$.
The observation angles $\left(\alpha,\beta\right)$ are then defined
in terms of the photon's local momentum as~\cite{Cunha:2018acu},
\begin{equation}
\sin\alpha=\frac{p^{\left(\theta\right)}}{p^{\left(t\right)}},\quad\tan\beta=\frac{p^{\left(\varphi\right)}}{p^{\left(r\right)}}.\label{eq:alpha-beta}
\end{equation}
Imposing the Hamiltonian constraint $\mathcal{H}=0$ and normalizing
the local momentum to $p^{\left(t\right)}=1$ uniquely determines
the initial four-momentum $p^{\mu}\left(x_{o}\right)$ from the angles
$\alpha$ and $\beta$. Finally, the coordinates on the observer\textquoteright s
image plane are defined for consistency as
\begin{equation}
x\equiv-r_{o}\beta,\quad y\equiv r_{o}\alpha.\label{eq:=000020celestial=000020coordinates}
\end{equation}

Following the thin disk model built by Gralla et al.~\cite{Gralla:2020srx},
the observed intensity at a given image position $\left(\alpha,\beta\right)$
is
\begin{equation}
I_{o}\left(\alpha,\beta\right)=\sum_{m=1}g_{s}^{4}\left(r_{m}\left(\alpha,\beta\right)\right)j_{e}\left(r_{m}\left(\alpha,\beta\right)\right),
\end{equation}
where $r_{m}$ denotes the radius of $m$th intersection of the backward-traced
geodesic with the disk, and $g_s$ is the redshift factor between the
observed and emitted photon frequencies. For a photon with momentum
$p^{\mu}$, the redshift factor is given by $g_s=p_{\mu}u_{o}^{\mu}/p_{\nu}u_{e}^{\nu}$,
where $u_{o}^{\mu}$ and $u_{e}^{\nu}$ are the four-velocities of
the observer and the emitter, respectively. The emission profile is
chosen as
\begin{equation}
j_{e}\left(r\right)=\frac{e^{-\left[\gamma+\textrm{arcsinh}\left(r-\mu/\sigma\right)\right]^{2}/2}}{\sqrt{\left(r-\mu\right)^{2}+\sigma^{2}}},
\end{equation}
with parameters $\gamma=-2$, $\mu=r_{\text{ISCO}}-M/3$ and $\sigma=M/4$,
which primarily concentrates emission near the ISCO. 

Moreover, we assume a prograde thin disk: matter outside the ISCO
moves along timelike circular geodesics in the equatorial plane, while
matter inside the ISCO plunges into the black hole with the same specific
energy and angular momentum as at the ISCO. For black holes possessing
a stable orbit island, such plunging matter may not populate the inner stable island. In such cases, we truncate
the disk at the outer boundary of the island. Under our observational
setup, this truncation has negligible impact on the resulting images.

\section{Images of Scalarized KN Black Holes}

\label{sec:Lensing-Images}

In this section, we begin with studying the behavior of the critical curve
for scalarized KN black holes within the slow-rotation approximation.
We then present numerically simulated images of
scalarized KN black holes and discuss the physical mechanisms that
produce the characteristic features associated with different types
of black hole. For numerical simulations, we set the observer
distance to $r_{o}=50M$, adopt the unit $M=1$ and introduce the
dimensionless charge $q\equiv Q/M$ and spin $\chi\equiv a/M$. The
observer's field of view is defined by the angular range $-\pi/30<\alpha,\beta<\pi/30$,
which is sampled with a resolution of $1000\times1000$ pixels. 

\subsection{Insight From Slow-rotation Approximation}

To gain analytic insight into how rotation modifies the critical curve
structure, we first consider scalarized KN black holes in the slow-rotation
approximation. In this regime, the metric functions can be expanded
as 
\begin{equation}
F_{i}\left(r,\theta\right)=F_{i}^{\left(0\right)}\left(r\right)+\mathcal{O}\left(\chi^{2}\right),\quad W\left(r,\theta\right)=W^{\left(1\right)}\left(r\right)\chi+\mathcal{O}\left(\chi^{3}\right),
\end{equation}
with $F_{1}^{\left(0\right)}\left(r\right)=F_{2}^{\left(0\right)}\left(r\right)$,
as required by the spherically symmetric limit. In practice, we take
$F_{i}^{\left(0\right)}\left(r\right)=F_{i}\left(r\right)$ directly
from the corresponding spherically symmetric numerical solution, and
extract
\begin{equation}
W^{\left(1\right)}\left(r\right)=\frac{1}{2\chi}\int_{0}^{\pi}W\left(r,\theta\right)\sin\theta d\theta,
\end{equation}
from the slowly rotating numerical solution.

At linear order in the dimensionless spin $\chi$, we assume that
the Hamilton--Jacobi action is separable and takes the form $S_{\text{slow}}=-Et+L_{z}\varphi+S_{r}\left(r\right)+S_{\theta}\left(\theta\right). $
Substituting $p_{\mu}=\partial S_{\text{slow}}/\partial x^{\mu}$
into the null condition gives the separated equations
\begin{align}
\frac{r^{4}e^{4F_{1}^{\left(0\right)}}}{N}\dot{r}^{2}+R\left(r\right) & =-\eta,\nonumber \\
r^{4}e^{4F_{1}^{\left(0\right)}}\dot{\theta}^{2}+\frac{b^{2}}{\sin^{2}\theta} & =\eta,
\end{align}
where
\begin{equation}
R\left(r\right)=-r^{2}\frac{e^{2F_{1}^{\left(0\right)}-2F_{0}^{\left(0\right)}}}{N}+2bW^{\left(1\right)}\frac{e^{2F_{1}^{\left(0\right)}-2F_{0}^{\left(0\right)}}}{N}\chi,
\end{equation}
is the radial potential, and $\eta$ denotes the energy-scaled Carter
constant.

The critical curve on the observer's image plane is determined by
unstable circular null orbits, which in the present approximation
reduce to photon spheres of the effective radial potential. Their
radii satisfy
\begin{equation}
R\left(r_{c}\right)=R^{'}\left(r_{c}\right)=0,\label{eq:=000020photon=000020sphere=000020condition}
\end{equation}
supplemented by the unstable condition $R^{''}\left(r_{c}\right)<0$,
where primes denote differentiation with respect to $r$. We further
expand the photon sphere radius and the corresponding critical Carter
constant as
\begin{equation}
r_{c}=r_{c}^{\left(0\right)}+r_{c}^{\left(1\right)}\chi+\mathcal{O}\left(\chi^{2}\right),\quad\eta_{c}=\eta_{c}^{\left(0\right)}+\eta_{c}^{\left(1\right)}\chi+\mathcal{O}\left(\chi^{2}\right).
\end{equation}
The zeroth-order quantities are determined by the non-rotating background,
while the first-order coefficients encode the leading rotational corrections.
Solving $\left(\ref{eq:=000020photon=000020sphere=000020condition}\right)$
to linear order in $\chi$, we obtain
\begin{align}
r_{c}^{\left(1\right)}= & \frac{2b\left(W^{\left(1\right)}\frac{e^{2F_{1}^{\left(0\right)}-2F_{0}^{\left(0\right)}}}{N}\right)_{r=r_{c}^{\left(0\right)}}^{'}}{\left(r^{2}\frac{e^{2F_{1}^{\left(0\right)}-2F_{0}^{\left(0\right)}}}{N}\right)_{r=r_{c}^{\left(0\right)}}^{''}},\label{eq:=000020first-order=000020sol}\\
\eta_{c}^{\left(1\right)}= & 2r_{c}^{\left(0\right)}r_{c}^{\left(1\right)}\left.\frac{e^{2F_{1}^{\left(0\right)}-2F_{0}^{\left(0\right)}}}{N}\right|_{r=r_{c}^{\left(0\right)}}-2b\left.W^{\left(1\right)}\frac{e^{2F_{1}^{\left(0\right)}-2F_{0}^{\left(0\right)}}}{N}\right|_{r=r_{c}^{\left(0\right)}}+r_{c}^{\left(0\right)2}r_{c}^{\left(1\right)}\left(\frac{e^{2F_{1}^{\left(0\right)}-2F_{0}^{\left(0\right)}}}{N}\right)_{r=r_{c}^{\left(0\right)}}^{'}.\nonumber 
\end{align}
Substituting the obtained critical parameters into $\left(\ref{eq:alpha-beta}\right)$,
one obtains the corresponding critical curve on the observer's image
plane.
\begin{figure}
\includegraphics[width=0.33\textwidth]{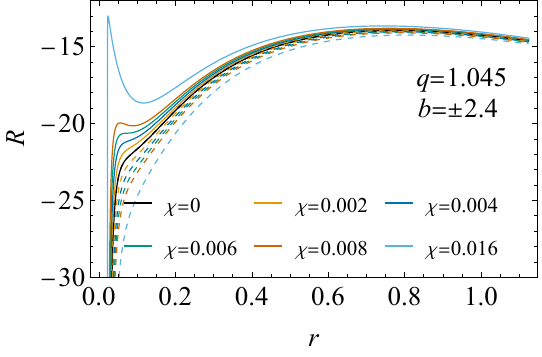}~\includegraphics[width=0.33\textwidth]{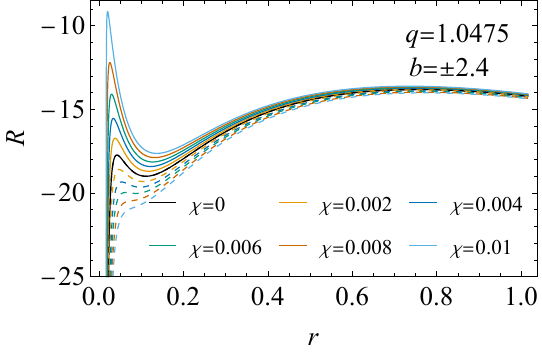}~\includegraphics[width=0.33\textwidth]{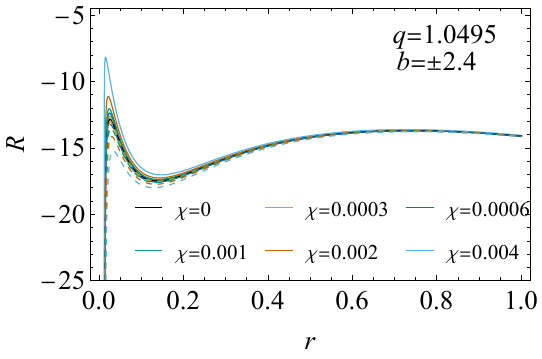}
\caption{Radial effective potential in the slow-rotation approximation for
three representative classes of scalarized black holes, shown for
several values of the spin $\chi$. The black solid line denotes the
non-rotating, spherically symmetric case $\left(\chi=0\right)$, while
the colored solid and dashed lines represent prograde $\left(b=2.4\right)$
and retrograde $\left(b=-2.4\right)$ photon potentials, respectively.
\textbf{Left:} black hole with a single peak in the non-rotating case,
where an inner peak gradually develops for prograde photons as $\chi$
increases, corresponding to a transition from Type I to Type II1,
and then Type II2. \textbf{Middle:} black hole with the inner peak
lower than the outer one in the non-rotating case; rotation enhances
the inner peak for prograde photons but suppresses it for retrograde
photons, allowing the black hole type to transition from Type III1
to Type III2 or Type II1, and eventually Type II2. \textbf{Right:}
black hole with the inner peak higher than the outer one in the non-rotating
case; with increasing $\chi$ , the prograde inner peak continues
to rise, while the retrograde one decreases, leading to a transition
from Type III3 to Type III2.}

\label{Fig:=000020radial=000020potential}
\end{figure}
\begin{figure}
\includegraphics[width=0.75\textwidth]{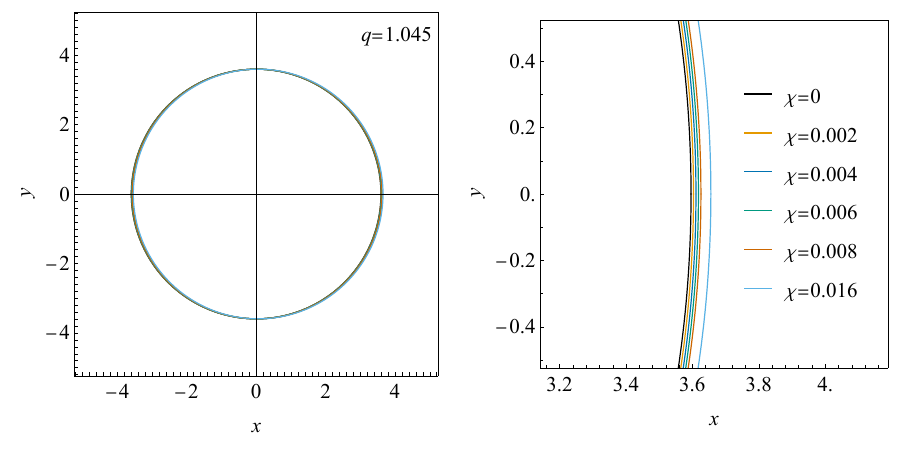}\caption{Critical curves for several values of the spin $\chi$ for scalarized
black hole with $q=1.045$, obtained from the first order solution
in $\left(\ref{eq:=000020first-order=000020sol}\right)$. Since
the outer peak of the effective potential varies only slightly, the
corresponding critical curve undergoes only a minor shift to the right,
as shown more clearly in the enlarged panel on the right. The critical
curve associated with the inner peak is not shown, as it is not captured
by the first order solution.}

\label{Fig:=000020critical-I}
\end{figure}
\begin{figure}
\includegraphics[width=0.95\textwidth]{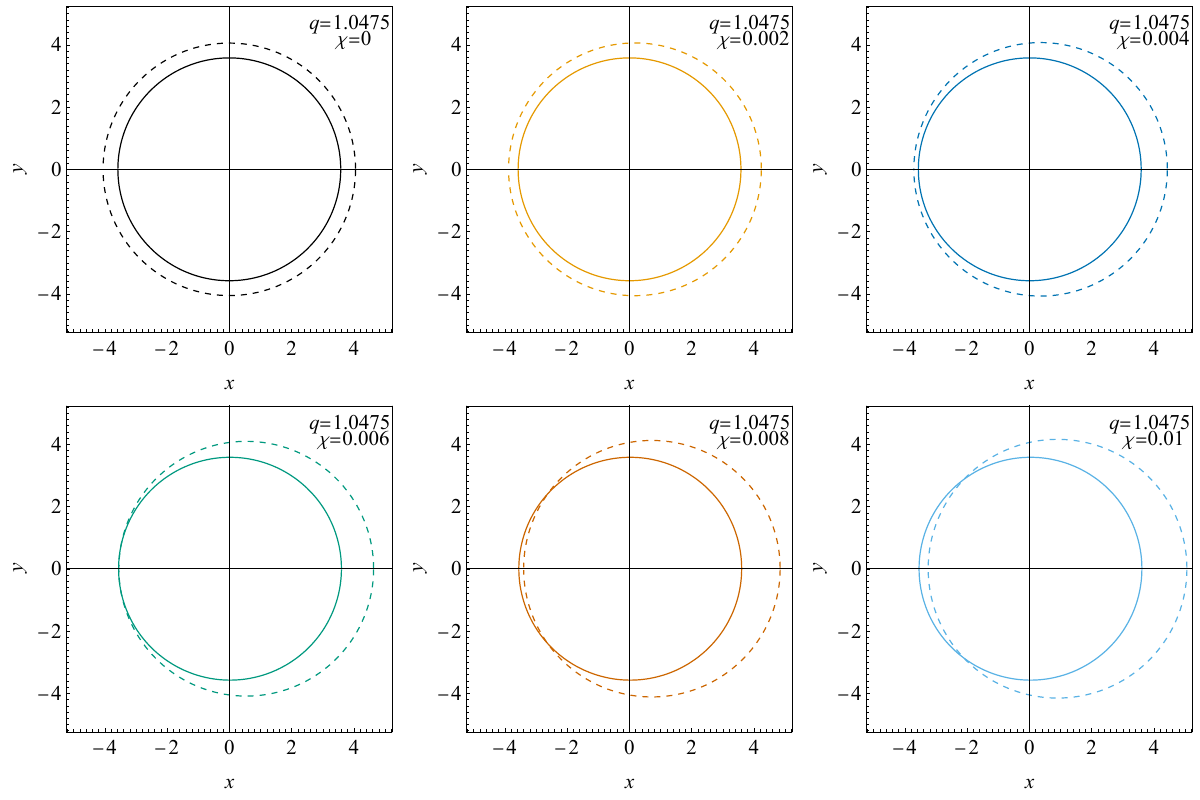}\caption{Critical curves for several values of spin $\chi$ for scalarized
black hole with $q=1.0475$, obtained from the first order solution
in $\left(\ref{eq:=000020first-order=000020sol}\right)$. The
solid and dashed curves denote the critical curves corresponding to
outer and inner peaks, respectively. As the spin $\chi$ increases,
the solid critical curve undergoes only a minor shift to the right.
However, the dashed critical curve significantly shifts to the right
due to the great change in the inner peak of the radial potential.
During this process, the critical curve associated with the inner
peak, which is initially invisible because it lies entirely outside
the critical curve of the outer peak, shifts to the right and becomes
partially visible once it intersects the outer critical curve.}

\label{Fig:=000020critical-II}
\end{figure}
\begin{figure}
\includegraphics[width=0.95\textwidth]{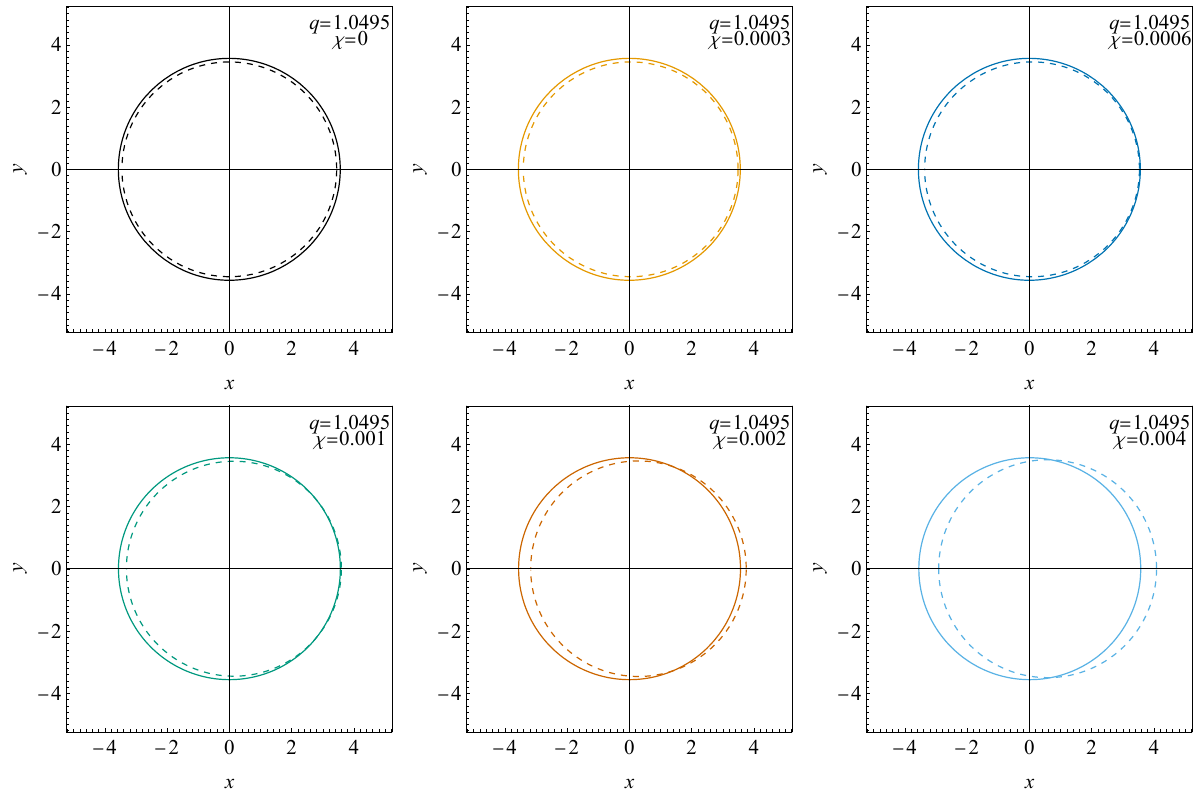}\caption{Critical curves for several values of spin $\chi$ for scalarized
black hole with $q=1.0495$, obtained from the first order solution
in Eqs. $\left(\ref{eq:=000020first-order=000020sol}\right)$. The
solid and dashed curves denote the critical curves corresponding to
outer and inner peaks, respectively. As
the spin  $\chi$ increases, the dashed critical curve, which is initially fully visible and entirely enclosed by the outer critical curve, shifts rightward and becomes only partially visible once it intersects the outer critical curve.}

\label{Fig:=000020critical-III}
\end{figure}

Figure~\ref{Fig:=000020radial=000020potential} shows the radial potential
for three representative classes of scalarized black holes within
the slow-rotation expansion. Photon motion is allowed only in the
region $-\eta>R\left(r\right)$. The black solid curves denote the
spherically symmetric cases, while the colored solid and dashed curves
correspond to prograde photons with $b=2.4$ and retrograde photons
with $b=-2.4$, respectively. The three panels represent, from left
to right, a black hole with a single-peak radial potential in the
non-rotating limit, a black hole with a double-peak radial potential
whose inner peak is lower than the outer one, and a black hole with
a double-peak radial potential whose inner peak is higher than the
outer one.

We first consider black holes whose radial potentials are shown in
the left panel of Fig.~\ref{Fig:=000020radial=000020potential}, with
the corresponding critical curves obtained from the first order solution
displayed in Fig.~\ref{Fig:=000020critical-I}. In the spherically
symmetric case, the black hole with a single-peak potential belongs
to Type I in the classification shown in Table~\ref{Table:=000020classification}
and gives rise to a single critical curve. As the spin $\chi$ increases,
the outer peak changes mildly, showing a slight increase for prograde
photons and a slight decrease for retrograde photons. This implies
that the critical curve undergoes only a small rightward shift as
the spin increases, as illustrated more clearly in Fig.~\ref{Fig:=000020critical-I}.
Meanwhile, for prograde photons, a new peak gradually forms at smaller
$r$. This indicates that the black hole transitions from Type I to
Type II1. With a further increase in $\chi$, the inner peak continues
to rise and eventually becomes higher than the outer peak. The black
hole then transitions to Type II2, in which some critical photons
associated with the inner peak are able to escape to infinity. This
suggests that an additional inner critical curve may appear on the
left side of the black hole image. However, this additional structure
is not captured by the first order solution, since the inner peak
is absent in the spherically symmetric limit.

We next consider black holes whose radial potentials are shown in
the middle panel of Fig.~\ref{Fig:=000020radial=000020potential},
with the corresponding critical curves obtained from the first order
solution displayed in Fig.~\ref{Fig:=000020critical-II}. In the spherically
symmetric case, the black hole with a double-peak potential, whose
inner peak is lower than the outer peak, belongs to Type III1 and
gives rise to a single critical curve associated with the outer peak.
As the spin $\chi$ increases, the outer peak remains nearly unchanged,
indicating only a mild variation of the outer critical curve, as shown
by the solid curves in Fig.~\ref{Fig:=000020critical-II}. In contrast,
the inner peak shows a pronounced dependence on the photon orientation:
for prograde photons, the inner peak increases with $\chi$, whereas
for retrograde photons it decreases and may even disappear. This indicates
that the black hole transitions from Type III1 either to Type III2
or to Type II1, depending on whether the inner peak for prograde photons
exceeds the outer peak before the inner peak for retrograde photons
vanishes, or vice versa. With a further increase in $\chi$, the configuration
eventually transitions to Type II2. As in the previous case, once
the inner peak becomes higher than the outer peak for prograde photons,
an additional inner critical curve may appear on the left side of
the black hole image. In this case, the appearance of the additional
inner critical curve is already captured by the first order solution,
as shown by the dashed curves in Fig.~\ref{Fig:=000020critical-II}.
Initially, the dashed critical curve lies outside the solid critical
curve and is therefore invisible. As $\chi$ increases, the dashed
critical curve partially enters the solid critical curve on the left
side of the image and becomes visible.

We finally consider black holes whose radial potentials are shown
in the right panel of Fig.~\ref{Fig:=000020radial=000020potential},
with the corresponding critical curves obtained from the first-order
solution displayed in Fig.~\ref{Fig:=000020critical-III}. In the
spherically symmetric case, the black hole with a double-peak potential,
whose inner peak is higher than the outer peak, belongs to Type III3
and gives rise to two critical curves, as shown in the first panel
of Fig.~\ref{Fig:=000020critical-III}. As the spin $\chi$ increases,
the outer peak remains nearly unchanged, and the inner peak shows
a strong dependence on the photon orientation: for prograde photons,
it continues to increase with $\chi$, whereas for retrograde photons
it decreases and may eventually become lower than the outer peak.
This indicates that the black hole transitions from Type III3 to Type
III2. Once the inner peak becomes lower than the outer peak for retrograde
photons, the inner critical curve may disappear on the right side
of the black hole image. This behavior is illustrated in Fig.~\ref{Fig:=000020critical-III},
where the dashed critical curve partially moves outside the solid
critical curve on the right side of the image as $\chi$ increases
and becomes partially invisible.

Overall, the appearance or disappearance of additional inner critical
curves can be understood directly from the way slow rotation modifies
the radial potential, especially the relative heights of its local
maxima. In the spherically symmetric limit, the effective potential
may contain two peaks associated with inner and outer unstable photon
spheres, but the visibility of the inner critical curve is determined
by whether the corresponding inner peak is higher than the outer one.
Once rotation is introduced, frame dragging breaks the symmetry between
prograde and retrograde photons. For prograde photons, the inner peak
increases and may eventually exceed the outer one, thereby giving
rise to a visible inner critical curve. For retrograde photons, by
contrast, the inner peak is suppressed and may even disappear, leading
to the partial disappearance of the inner critical curve. This spin-induced
asymmetry may cause the inner critical curve to appear only partially.
We emphasize, however, that the slow-rotation approximation is intended
only as a qualitative guide. Near the inner photon sphere, higher-order
rotational corrections may become important. Although the present
analysis is useful for understanding the overall behavior of the critical
curves, a quantitatively reliable determination of their positions
and shapes requires full ray-tracing simulations presented in the
next subsection.

\subsection{Numerical Results}

\begin{figure}
\includegraphics[width=0.25\textwidth]{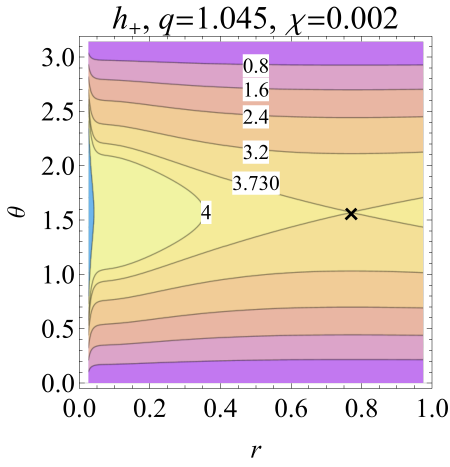}~\includegraphics[width=0.25\textwidth]{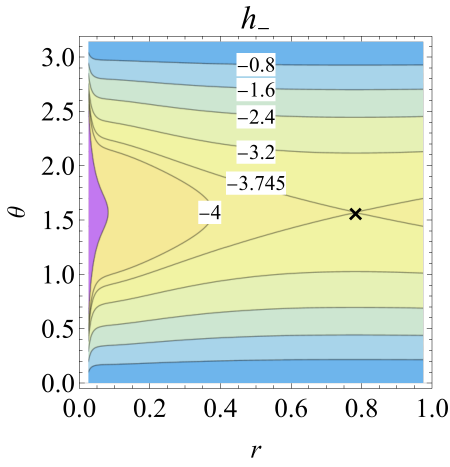}~\includegraphics[width=0.25\textwidth]{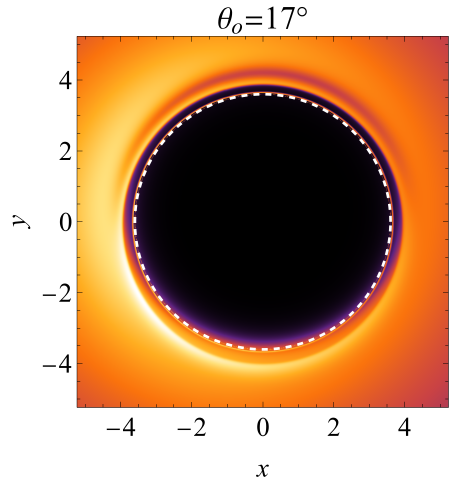}~\includegraphics[width=0.25\textwidth]{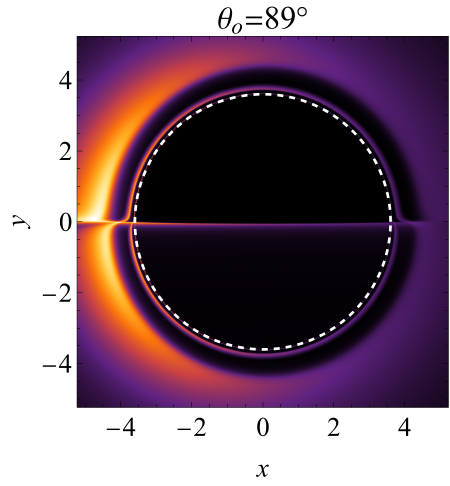}

\caption{Plots for a scalarized KN black hole of Type I ($q=1.045$ and $\chi=0.002$),
admitting one prograde and one retrograde unstable light ring. The
first and second columns show the effective potential contours for
prograde and retrograde photons, respectively. Crosses mark the saddle
points corresponding to unstable light rings. The prograde light ring
appears at $r_{c+}=0.770$ with a critical impact parameter $b_{c+}=3.730$,
while the retrograde light ring appears at $r_{c-}=0.783$ with $b_{c-}=-3.745$.
The third and fourth columns display the observed accretion disk images
at inclination angles of $17^{\circ}$ and $89^{\circ}$, respectively.
The dashed lines denote the critical curves on the image plane, traced
by photons at the threshold between escaping and falling into the
photon shell. Due to the small spin, the images closely resemble those
of a spherically symmetric black hole, featuring a bright ring surrounding
a dark central region. At higher inclinations, the Doppler effect
from the rotating accretion disk produces a left-bright/right-dark
asymmetry.}
\label{Fig:=000020image-I}
\end{figure}

We now present full ray-tracing results and compare them with the
qualitative picture inferred from the slow-rotation analysis. We begin
by examining a scalarized KN black hole with $q=1.045$ and $\chi=0.002$.
The first and second panels of Fig.~\ref{Fig:=000020image-I} depict
the effective potentials of the prograde and retrograde photons, respectively.
Each potential exhibits a saddle point, located at $r_{c+}=0.770$
and $r_{c-}=0.783$, corresponding to the prograde and retrograde
unstable light rings. This identifies the solution as a Type I black
hole. When the photon impact parameter reaches the critical values
$b_{c+}=3.730$ and $b_{c-}=-3.745$, the allowed regions disconnect
at the light rings. As a result, incident photons with $b>b_{c+}$
or $b<b_{c-}$ cannot penetrate into the region interior to the corresponding
light ring. However, this does not imply that all photons with $b<b_{c+}$
or $b>b_{c-}$ can always enter the region inside the corresponding
light ring. For off-equatorial photons, the relevant behavior is instead
governed by the photon shell formed by bound null geodesics between
the prograde and retrograde light rings as in the Kerr spacetime.

The third and fourth panels of Fig.~\ref{Fig:=000020image-I} indeed
show Kerr-like images with a single critical curve, traced by the
dashed line, and with identifiable higher-order photon rings accumulating
toward it. This behavior is fully consistent with the slow-rotation
analysis of the single-peak case. There, the outer peak was shown
to vary only weakly with $\chi$, implying that the associated critical
curve should undergo only a small deformation. At low inclination
$\left(17^{\circ}\right)$, the bright ring extends inside the critical
curve due to direct emission from the disk lying inside the photon
shell, whose boundary defines the inner shadow~\cite{Chael:2021rjo}.
At high inclination $\left(89^{\circ}\right)$, the Doppler effect
produces a pronounced left-bright/right-dark asymmetry. 

\begin{figure}
\includegraphics[width=0.23\textwidth]{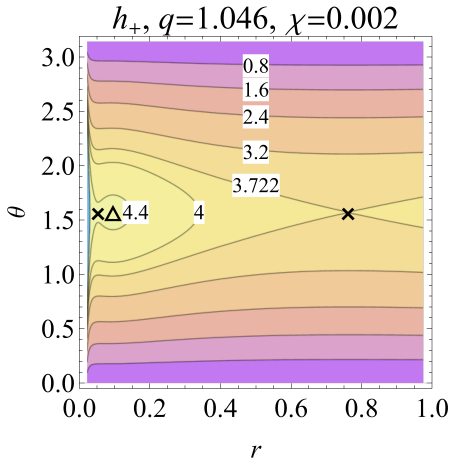}~\includegraphics[width=0.23\textwidth]{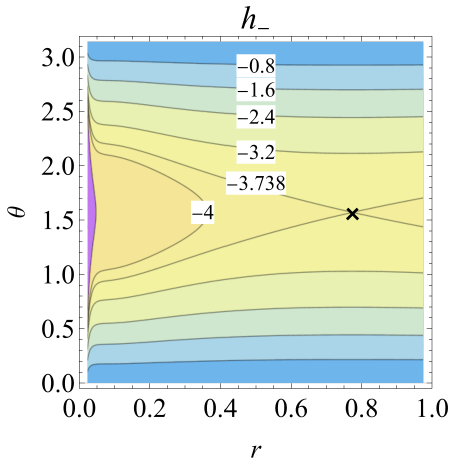}~\includegraphics[width=0.23\textwidth]{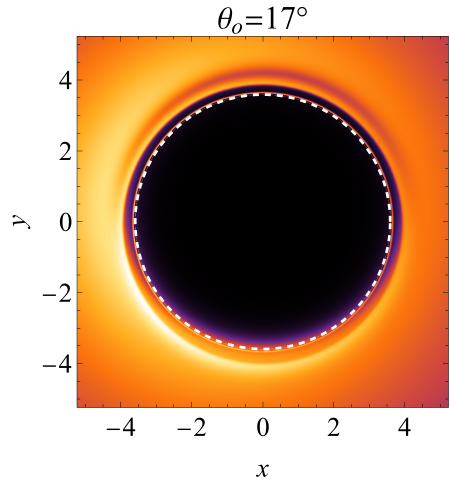}~\includegraphics[width=0.23\textwidth]{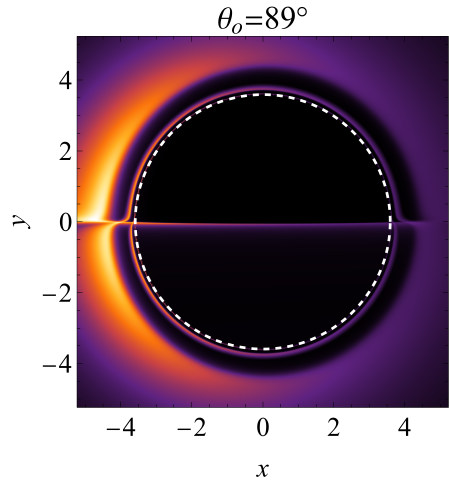}

\includegraphics[width=0.23\textwidth]{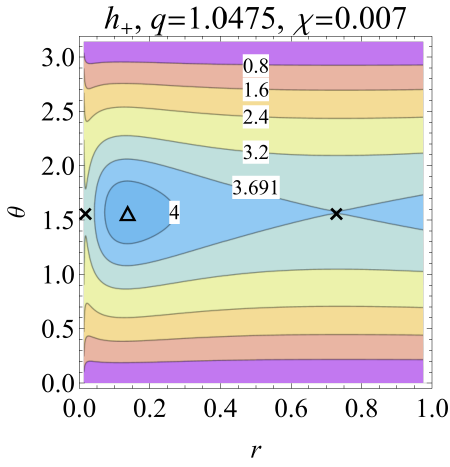}~\includegraphics[width=0.23\textwidth]{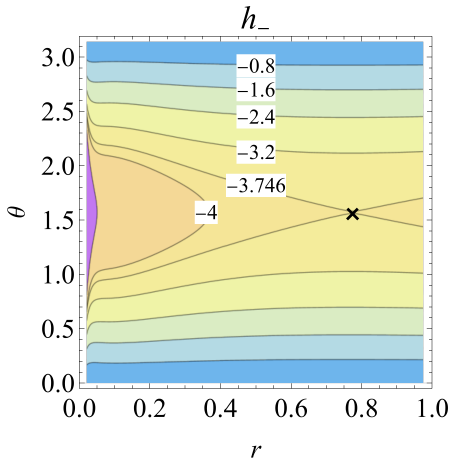}~\includegraphics[width=0.23\textwidth]{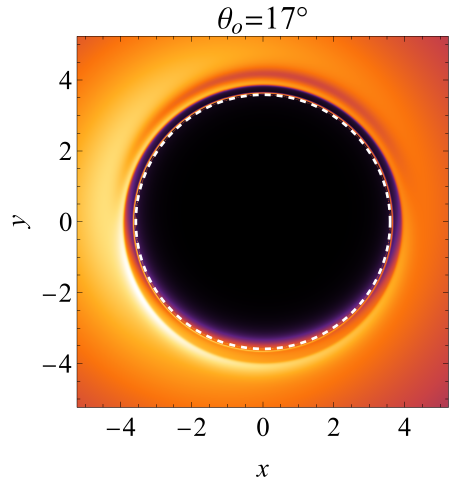}~\includegraphics[width=0.23\textwidth]{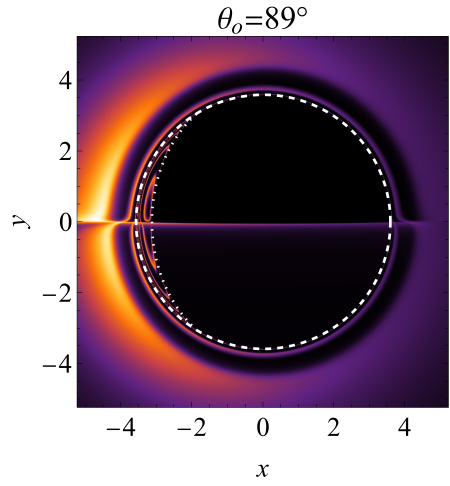}

\caption{Plots for scalarized KN black holes of Type II1 (Upper: $q=1.046$
and $\chi=0.002$) and Type II2 (Bottom: $q=1.0475$ and $\chi=0.007$),
admitting two prograde and one retrograde unstable light rings. The
triangles mark the maxima corresponding to stable light rings. For
the Type II1 black hole, the prograde unstable light rings appear
at $r_{c+,\text{in}}=0.053$ and $r_{c+,\text{out}}=0.762$ with critical
impact parameters $b_{c+,\text{in}}=4.419$ and $b_{c+,\text{out}}=3.722$,
respectively, while the retrograde light ring appears at $r_{c-}=0.774$
with $b_{c-}=-3.738$. Although this type of black hole possesses
an additional unstable light ring, the simulated observed images remain
similar to those of the Type I black hole. For the Type II2 black
hole, the prograde unstable light rings appear at $r_{c+,\text{in}}=0.017$
and $r_{c+,\text{out}}=0.729$ with critical impact parameters $b_{c+,\text{in}}=3.252$
and $b_{c+,\text{out}}=3.691$, respectively; the retrograde light
ring appears at $r_{c-}=0.774$ with $b_{c-}=-3.746$. At a low inclination
angle of $17^{\circ}$, the observed images remain similar to those
of the Type I black hole. However, at a high inclination  of $89^{\circ}$,
an additional partial inner critical curve  emerges, marked by a dotted
line. This inner critical curve intersects the outer critical curve
(dashed line), and the region enclosed by them gives rise to new higher-order
images.}
\label{Fig:=000020image-II}
\end{figure}

We next consider Type II scalarized KN black holes, whose effective
potentials admit one prograde stable light ring, two prograde unstable
light rings, and one retrograde unstable light ring. The upper row
of Fig.~\ref{Fig:=000020image-II} corresponds to a Type II1 black
hole with $q=1.046$ and $\chi=0.002$. The first two panels show
that the prograde and retrograde effective potentials exhibit three
saddle points and one local maximum on the equatorial plane. The saddle
points at $r_{c+,\text{in}}=0.053$ and $r_{c+,\text{out}}=0.762$
with associated critical impact parameters $b_{c+,\text{in}}=4.419$
and $b_{c+,\text{out}}=3.722$, correspond to the inner and outer
prograde unstable light rings. The saddle point at $r_{c-}=0.774$
with $b_{c-}=-3.738$ corresponds to the retrograde unstable light
ring. The intervening local maximum (marked by a triangle) identifies
the prograde stable light ring.

It is noteworthy that the impact parameter of the inner prograde unstable
light ring is larger than that of the outer one. Therefore, near-critical
photons generated by the inner prograde unstable light ring cannot
propagate through the outer prograde unstable light ring and therefore
cannot be detected by a distant observer. Consequently, the resulting
accretion disk images closely resemble those of the Type I black hole,
as shown in the third and fourth panels. The photon shell between
the outer prograde and retrograde unstable light rings again generates
the critical curve. This suggests that, for this specific parameter
set, the additional inner light ring does not significantly affect
the observable shadow or the photon ring structure accessible to distant
observers.

The bottom row of Fig.~\ref{Fig:=000020image-II} shows the results
for a Type II2 black hole with $q=1.0475$ and $\chi=0.007$. The
effective potentials share a structure similar to Type II1. The inner
and outer prograde unstable light rings shift to $r_{c+,\text{in}}=0.017$
and $r_{c+,\text{out}}=0.729$ with $b_{c+,\text{in}}=3.252$ and
$b_{c+,\text{out}}=3.691$. The retrograde light ring is located at
$r_{c-}=0.774$ with $b_{c-}=-3.746$. 

In this case, the impact parameter of the inner prograde unstable
light ring is smaller than that of the outer one, allowing near-critical
photons from the inner ring to propagate outward and reach the observer.
At a low inclination of $17^{\circ}$, the observed image again closely
resembles those of Type I and Type II1 black holes. At a high inclination
of $89^{\circ}$, however, a new feature emerges: in addition to the
outer critical curve (dashed line), an inner critical curve (dotted
line) appears. The outer curve is generated by the photon shell between
the outer prograde and retrograde unstable light rings, while the
inner curve originates from an inner photon shell near the inner prograde
unstable light ring. The two critical curves intersect, and the region
enclosed between them exhibits new crescent-like higher-order images
generated by near-bound trajectories that travel between the outer
and inner photon shells. This rich lensing structure marks a strong
qualitative deviation from the Type I observational signatures. This
provides a clear numerical realization of the additional inner critical
curve anticipated in the slow-rotation approximation.

\begin{figure}
\includegraphics[width=0.23\textwidth]{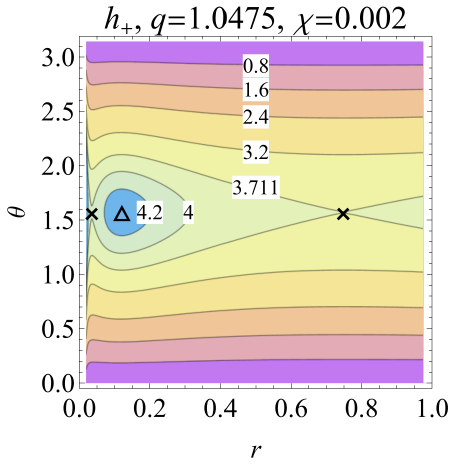}~\includegraphics[width=0.23\textwidth]{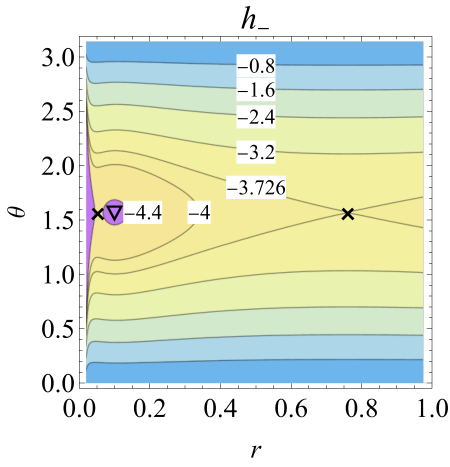}~\includegraphics[width=0.23\textwidth]{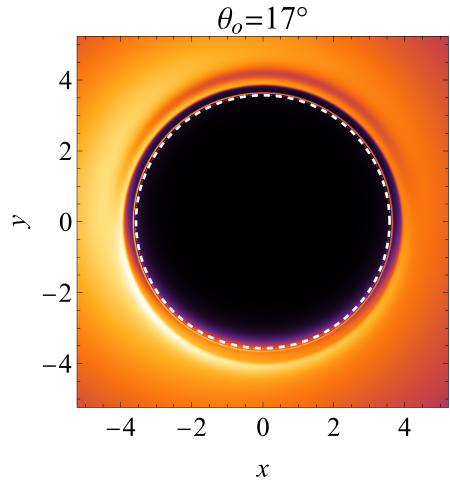}~\includegraphics[width=0.23\textwidth]{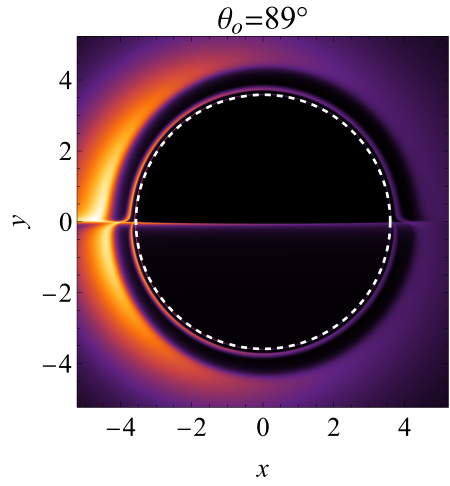}

\includegraphics[width=0.23\textwidth]{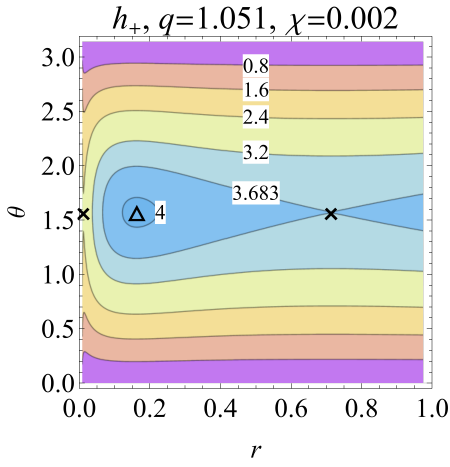}~\includegraphics[width=0.23\textwidth]{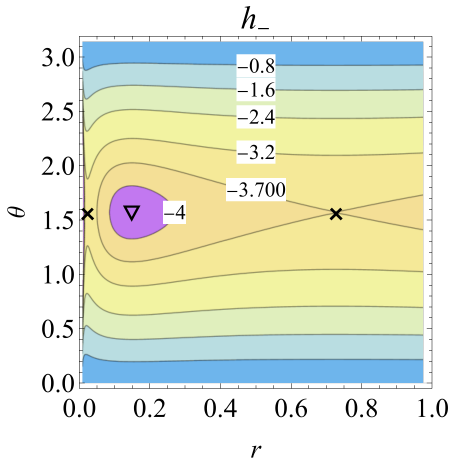}~\includegraphics[width=0.23\textwidth]{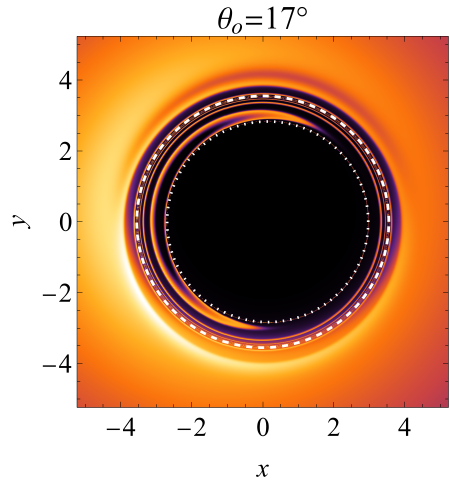}~\includegraphics[width=0.23\textwidth]{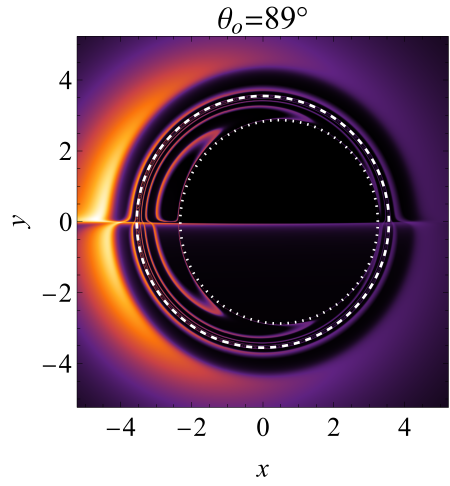}\caption{Plots for scalarized KN black holes of Type III1 (Upper: $q=1.0475$
and $\chi=0.002$) and Type III3 (Bottom: $q=1.051$ and $\chi=0.002$),
admitting two prograde and two retrograde unstable light rings. The
triangles mark the maxima corresponding to stable light rings. For
the Type III1 black hole, the prograde unstable light rings appear
at $r_{c+,\text{in}}=0.035$ and $r_{c+,\text{out}}=0.748$ with critical
impact parameters $b_{c+,\text{in}}=4.007$ and $b_{c+,\text{out}}=3.711$,
respectively, while the retrograde unstable light rings appear at
$r_{c-,\text{in}}=0.052$ and $r_{c-,\text{out}}=0.761$ with $b_{c-,\text{in}}=-4.383$
and $b_{c-,\text{out}}=-3.726$, respectively. Although this type
of black hole possesses two additional unstable light rings, the simulated
images remain similar to those of the Type I black hole. For the Type
III3 black hole, the prograde unstable light rings appear at $r_{c+,\text{in}}=0.011$
and $r_{c+,\text{out}}=0.713$ with critical impact parameters $b_{c+,\text{in}}=2.431$
and $b_{c+,\text{out}}=3.683$, respectively; the retrograde light
rings appear  at $r_{c-,\text{in}}=0.023$ and $r_{c-,\text{out}}=0.728$
with critical impact parameters $b_{c-,\text{in}}=-3.375$ and $b_{c-,\text{out}}=-3.700$,
respectively. An additional closed inner critical curve emerges inside
 the outer critical curve, marked by a dotted line, with several
additional higher-order images arising between them.}
\label{Fig:=000020image-III1/III3}
\end{figure}

We now turn to scalarized KN black holes of Type III, which admit
four unstable light rings (two prograde and two retrograde) together
with two stable light rings (one prograde and one retrograde).

The upper row of Fig.~\ref{Fig:=000020image-III1/III3} corresponds
to a Type III1 black hole with parameters $q=1.0475$ and $\chi=0.002$.
As shown in the first two panels, the prograde and retrograde effective
potentials each possess an inner and an outer saddle point on the
equatorial plane. The prograde unstable light rings are located at
$r_{c+,\text{in}}=0.035$ and $r_{c+,\text{out}}=0.748$ with $b_{c+,\text{in}}=4.007$
and $b_{c+,\text{out}}=3.711$. The retrograde unstable light rings
appear at $r_{c-,\text{in}}=0.052$ and $r_{c-,\text{out}}=0.761$
with $b_{c-,\text{in}}=-4.383$ and $b_{c-,\text{out}}=-3.726$. The
local maxima separating the inner and outer saddles correspond to
stable light rings. Despite this richer orbital structure, the observed
images, shown in the third and fourth panels, closely resemble those
of the Type I and Type II1 black holes. This is analogous to the case
of a spherically symmetric black hole with a double-peaked potential,
in which the inner peak is lower than the outer one. Here, near-bound
photons originating from the inner photon shell are obscured by the
outer photon shell and do not produce an observable critical curve
for a distant observer. This result is also consistent with the estimation
obtained in the slow-rotation approximation.

The bottom row of Fig.~\ref{Fig:=000020image-III1/III3} presents
the results for a Type III3 black hole with parameters $q=1.051$
and $\chi=0.002$. The unstable light rings are now located at $r_{c+,\text{in}}=0.011$
and $r_{c+,\text{out}}=0.713$ with critical impact parameters $b_{c+,\text{in}}=2.431$
and $b_{c+,\text{out}}=3.683$, and $r_{c-,\text{in}}=0.023$ and
$r_{c-,\text{out}}=0.728$ with $b_{c-,\text{in}}=-3.375$ and $b_{c-,\text{out}}=-3.700$.
In contrast to the Type III1 case, we now have $b_{c+,\text{in}}<b_{c+,\text{out}}$
and $b_{c-,\text{in}}>b_{c-,\text{out}}$. These changed relations
of the unstable light rings significantly modify the lensing behavior.
A closed inner critical curve (dotted) now appears entirely inside
the outer critical curve (dashed). The region between the two non-intersecting
curves contains multiple new higher-order images produced by near-bound
trajectories that become temporarily trapped within the two photon
shells. This behavior is analogous to that of a black hole with a
double-peaked potential, in which the inner peak is higher than the
outer one. Besides, the inner critical curve is clearly shifted to
the right of the center, which is also consistent with the prediction
of the slow-rotation approximation.

\begin{figure}
\includegraphics[width=0.23\textwidth]{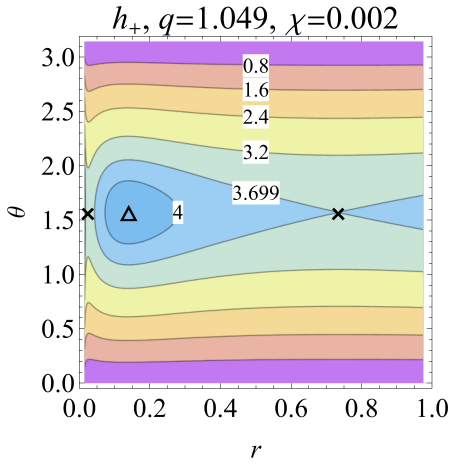}~\includegraphics[width=0.23\textwidth]{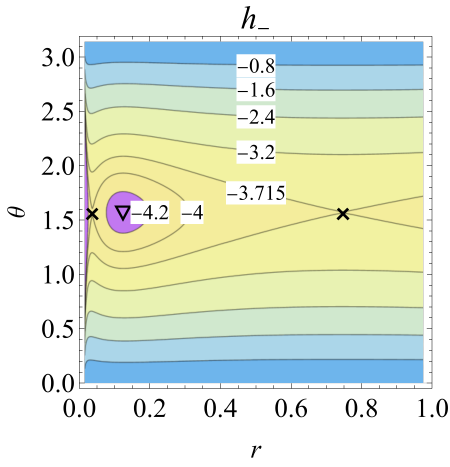}~\includegraphics[width=0.23\textwidth]{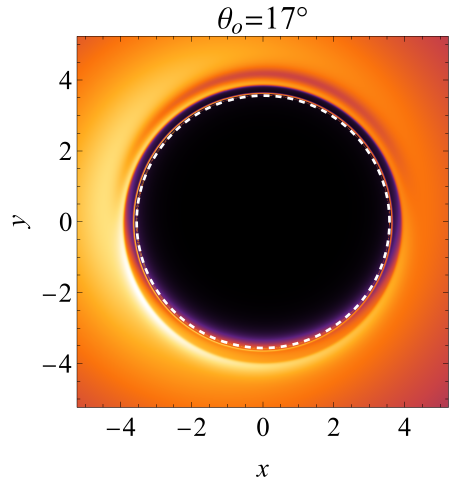}~\includegraphics[width=0.23\textwidth]{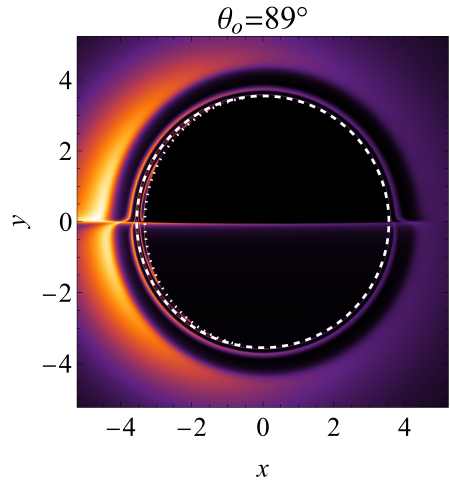}

\includegraphics[width=0.23\textwidth]{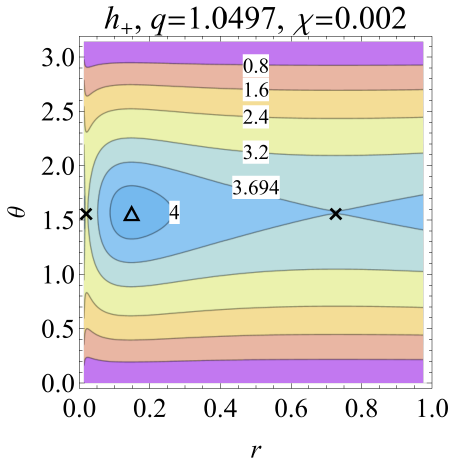}~\includegraphics[width=0.23\textwidth]{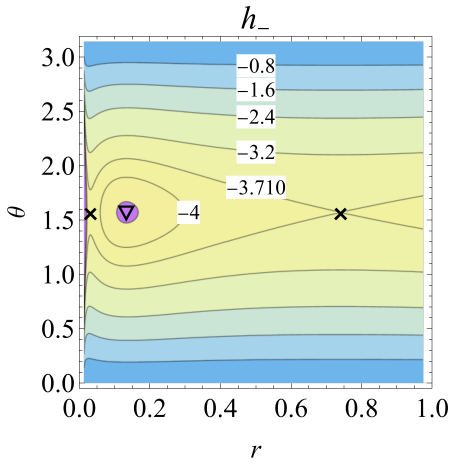}~\includegraphics[width=0.23\textwidth]{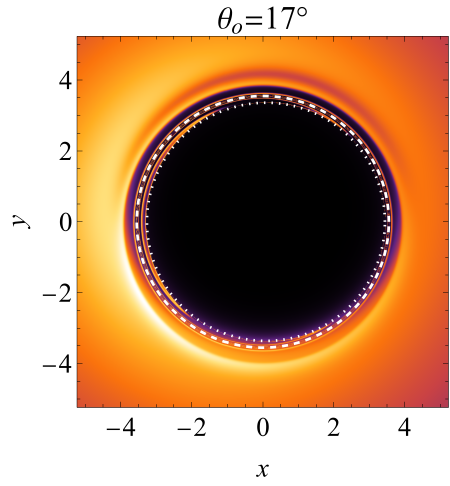}~\includegraphics[width=0.23\textwidth]{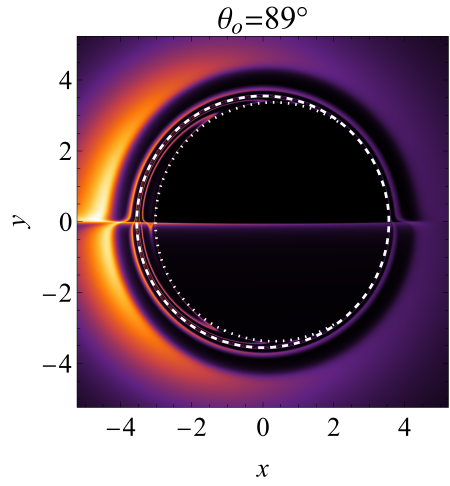}\caption{Plots for two scalarized KN black holes of Type III2, admitting two
prograde and two retrograde unstable light rings. The triangles mark
the maxima corresponding to stable light rings. \textbf{Upper: }$q=1.049$
and $\chi=0.002$. In this case, the prograde unstable light rings
appear at $r_{c+,\text{in}}=0.024$ and $r_{c+,\text{out}}=0.734$
with critical impact parameters $b_{c+,\text{in}}=3.462$ and $b_{c+,\text{out}}=3.699$,
respectively, while the retrograde unstable light rings appear at
$r_{c-,\text{in}}=0.037$ and $r_{c-,\text{out}}=0.747$ with $b_{c-,\text{in}}=-4.003$
and $b_{c-,\text{out}}=-3.715$, respectively. An additional partial
inner critical curve emerges inside the outer critical curve at a
high inclination  of $89^{\circ}$. \textbf{Bottom:} $q=1.0497$ and
$\chi=0.002$. The prograde unstable light rings appear at $r_{c+,\text{in}}=0.019$
and $r_{c+,\text{out}}=0.727$ with critical impact parameters $b_{c+,\text{in}}=3.152$
and $b_{c+,\text{out}}=3.694$, respectively; the retrograde light
rings appear at $r_{c-,\text{in}}=0.031$ and $r_{c-,\text{out}}=0.740$
with critical impact parameters $b_{c-,\text{in}}=-3.797$ and $b_{c-,\text{out}}=-3.710$,
respectively. An additional closed inner critical curve emerges at
a low inclination angle of $17^{\circ}$, while the inner critical
curve becomes partially visible at a high inclination  of $89^{\circ}$.}
\label{Fig:=000020image-III2}
\end{figure}

As a further example, we consider two scalarized KN black holes of
Type III2. The upper row of Fig.~\ref{Fig:=000020image-III2} corresponds
to a Type III2 black hole with parameters $q=1.049$ and $\chi=0.002$.
The prograde unstable light rings appear at $r_{c+,\text{in}}=0.024$
and $r_{c+,\text{out}}=0.734$ with $b_{c+,\text{in}}=3.462$ and
$b_{c+,\text{out}}=3.699$. The retrograde unstable light rings appear
at $r_{c-,\text{in}}=0.037$ and $r_{c-,\text{out}}=0.747$ with $b_{c-,\text{in}}=-4.003$
and $b_{c-,\text{out}}=-3.715$. In this configuration, $b_{c+,\text{in}}<b_{c+,\text{out}}$,
whereas $b_{c-,\text{in}}<b_{c-,\text{out}}$, implying that near-bound
photons from the inner prograde light ring can penetrate the outer
photon shell, while those from the inner retrograde light ring cannot.
As a result, a partial inner critical curve and new higher-order images
appear on the left side of the image at a high inclination of $89^{\circ}$.
Conversely, no such features appear on the right side. This behavior
is consistent with the prediction for Type III2 black holes within
the slow-rotation approximation. At a low inclination of $17^{\circ}$,
no additional inner critical curve appears, since a substantial fraction
of the photon momentum is in the $\theta$-direction, leading to relatively
small $\left|b\right|$ and hence small rotational corrections. From
the perspective of the slow-rotation approximation, this black hole
therefore remains more closely related to the Type III1 black hole,
which likewise does not exhibit an inner critical curve.

The bottom row of Fig.~\ref{Fig:=000020image-III2} shows the results
for another Type III2 configuration with parameters $q=1.0497$ and
$\chi=0.002$. The prograde unstable light rings appear at $r_{c+,\text{in}}=0.019$
and $r_{c+,\text{out}}=0.727$ with $b_{c+,\text{in}}=3.152$ and
$b_{c+,\text{out}}=3.694$. The retrograde unstable light rings appear
at $r_{c-,\text{in}}=0.031$ and $r_{c-,\text{out}}=0.740$ with $b_{c-,\text{in}}=-3.797$
and $b_{c-,\text{out}}=-3.710$. In this configuration, the relations
$b_{c+,\text{in}}<b_{c+,\text{out}}$ and $b_{c-,\text{in}}<b_{c-,\text{out}}$
still hold. Therefore, at a high inclination of $89^{\circ}$, additional
higher-order images and a partial inner critical curve appear on the
left side of the image, while none are present on the right side.
In contrast to the previous Type III2 case, a closed inner critical
curve appears at a low inclination of $17^{\circ}$, indicating that
this black hole is more closely related to the Type III3 black hole.

\begin{figure}
\includegraphics[width=1\textwidth]{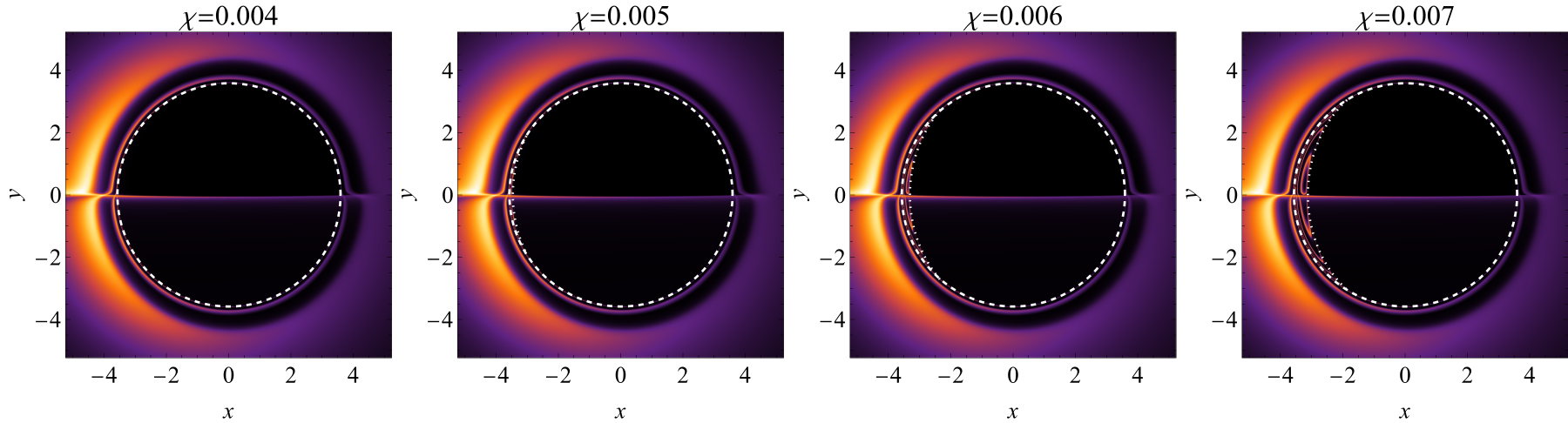}

\caption{Observed images of scalarized KN black holes with $q=1.0475$ at a
high inclination angle of $89^{\circ}$ for several values of the
spin, $\chi=0.004$, $0.005$, $0.006$ and $0.007$ (from left to right). The
dashed and dotted lines denote the outer and inner critical curves,
respectively. As the spin increases, the outer critical curve undergoes
only a mild deformation, while an additional inner critical curve
emerges on the left side of the image and becomes progressively more
visible. The region between the two curves gives rise to new higher-order
images, produced by near-bound photons.}

\label{Fig:=000020figchivary}
\end{figure}

We have focused on the slowly rotating case and presented a rich set of behaviors from both analytical and numerical perspectives. Before concluding this section, we turn to the effect of increasing spin on the black hole images. In particular, we focus on how the inner critical curve gradually emerges as $\chi$ increases. Figure~\ref{Fig:=000020figchivary} further illustrates the
spin dependence of the images for scalarized KN black holes with $q=1.0475$
and $\chi=0.004$, $0.005$, $0.006$ and $0.007$ at a high inclination of $89^{\circ}$.
At relatively low spin, the image is still dominated by the outer
photon shell and remains close to the Kerr black hole with a single
critical curve. As the spin increases, frame dragging enhances the
contribution of prograde inner near-bound photons associated with
the inner photon shell. Consequently, an additional inner critical
curve starts to appear on the left side of the image and becomes progressively
more prominent. This behavior is consistent with the slow-rotation
analysis discussed above: for the $q=1.0475$ branch, increasing spin
mainly drives the inner prograde near-bound photons toward visibility,
whereas the corresponding retrograde inner near-bound photons remain
suppressed. As a result, the additional inner critical curve is not
fully closed, but appears first as a partial arc on the left side
of the image. 

\section{Conclusions}

\label{sec:CONCLUSIONS}

\begin{table}
\begin{tabular}{ccc}
\hline 
\multirow{2}{*}{Type} & \multirow{2}{*}{High inclination} & \multirow{2}{*}{Low inclination}\tabularnewline
 &  & \tabularnewline
\hline 
I & Absent & Absent\tabularnewline
\hline 
II1 & Absent & Absent\tabularnewline
II2 & Partial & Absent\tabularnewline
\hline 
III1 & Absent & Absent\tabularnewline
III2 & Partial & Parameter dependent\tabularnewline
III3 & Closed & Closed\tabularnewline
\hline 
\end{tabular}

\caption{Existence and closure of the inner critical curve for different types
of scalarized KN black holes at high and low inclinations. Type I, II1 and III1 black holes are Kerr-like cases, exhibiting only a single outer critical curve and no inner critical curve. Type II2 black holes exhibit a partially visible inner critical curve only at high inclination. For Type
III2 black holes, the inner critical curve is  partially visible at a high inclination, while at a low inclination it can be absent, partial, or closed, depending on the black hole parameters. Type III3 black holes always exhibit two closed critical curves.  In addition, higher-order images always appear between the inner and outer critical curves.}

\label{Table:=000020classification-critical=000020curve}
\end{table}

In this work, we have studied the optical appearance of slowly rotating
scalarized KN black holes in the EMS theory with exponential coupling.
Our analysis focused on the relation between light rings and the observable
properties of thin disk images. In particular, by studying the effective
potentials for prograde and retrograde photon motion, we clarified
how differences in the light ring structure among scalarized solutions
are reflected in the resulting lensing images.

Our numerical results are also in good agreement with the expectations
from the slow-rotation approximation. In particular, the slow-rotation
analysis correctly captures the qualitative relation between the black
hole type and the appearance or absence of additional inner critical
curves. It also provides a useful way to understand the transition
of image features between different scalarized solutions. In this
sense, the full ray-tracing results confirm that the slow-rotation
approximation already contains the essential physical mechanism responsible
for the main image patterns discussed in this work.

Our results show that, although some scalarized KN black holes can still display Kerr-like images dominated by the usual outer photon shell and a single critical curve, other scalarized solutions exhibit features that are absent in the Kerr case. Most importantly, the presence of multiple light rings implies the existence of additional inner photon shells outside the horizon. Consequently, scalarized KN black holes can develop additional inner critical curves inside the usual outer critical curve. Depending on the black hole parameters
and the observer inclination, these inner critical curves may be absent,
partial or closed, as summarized in Table~\ref{Table:=000020classification-critical=000020curve}.

Another striking departure from the Kerr case is the emergence of new higher-order images in the region between the outer and inner critical curves. These nested higher-order images arise from near-bound photon trajectories propagating between two photon shells. Some of them present crescent-like shapes, in clear contrast to the typically circular higher-order images found in the Kerr case.

These results indicate that scalarization can leave nontrivial observational
imprints on black hole images, even when the outer critical curve
remains close to the Kerr case. The appearance of additional inner
critical curves and extra higher-order images provides a potentially
distinctive signature of scalarized KN black holes in the strong field
regime. Such features may offer a useful probe of scalarized black
holes in future very long baseline interferometric observations with
sufficiently high angular resolution.

\begin{acknowledgments}
We are grateful to Guangzhou Guo for useful discussions and valuable
comments. This work is supported in part by NSFC (Grant No. 12275183, No. 12275184, No. 12525503, No. 11875196, No. 12588101 and No. 12447101). 
\end{acknowledgments}

\bibliographystyle{unsrturl}
\bibliography{ref}

\end{document}